\documentclass[notoc,paper,letterpaper]{JHEP3}
\usepackage{amssymb} 
\usepackage{graphicx}     		
\newcommand{\be}{\begin{equation}} 
\newcommand{\ee}{\end{equation}} 
\newcommand{\bea}{\begin{eqnarray}} 
\newcommand{\eea}{\end{eqnarray}}

\def\bar{\overline}
\def\til{\widetilde}

\def\op{{\oplus}}
\def\cd{{\cdot}}
\def\tf#1#2{{\textstyle{\frac{#1}{#2}}}}

\def\Z{\mathbb{Z}}

\def\U{{\rm u}} 			
\def\SU{{\rm su}} 		
\def\SO{{\rm so}} 		
\def\Sp{{\rm sp}} 		
\def\ff{{\mathfrak f}}
\def\tff{{\til\ff}}
\def\gf{{\mathfrak g}}
\def\tgf{{\til\gf}}
\def\hf{{\mathfrak h}}
\def\br{{\bf r}}
\def\tbr{{\til\br}}
\def\hra{\hookrightarrow}

\def\with{\mathop{\,\textstyle{\rm w/}\,}}
\def\SCFT{{\rm SCFT}}

\title{Infinite coupling duals of N=2 gauge theories\\
and new rank 1 superconformal field theories}
\author{Philip C. Argyres and John R. Wittig\\ 
Physics Department, University of Cincinnati, 
Cincinnati OH 45221-0011\\
\email{argyres@physics.uc.edu, jwittig0126@gmail.com}}

\abstract{We show that a proposed duality \cite{as0711} between 
infinitely coupled gauge theories and superconformal field theories 
(SCFTs) with weakly gauged flavor groups predicts the existence of new 
rank 1 SCFTs.  These superconformal fixed point theories have the same 
Coulomb branch singularities as the rank 1 $E_6$, $E_7$, and $E_8$ SCFTs, 
but have smaller flavor symmetry algebras and different central
charges.  Gauging various subalgebras of the flavor algebras of these 
rank 1 SCFTs provides many examples of infinite-coupling dualities, 
satisfying an intricate set of consistency checks.  They also
provide examples of $N=2$ conformal theories with marginal
couplings but no weak-coupling limits.}

\begin{document}


A recently proposed expansion of the notion of S-duality \cite{as0711} 
provides a new tool for exploring the set of $N=2$ 
SCFTs by identifying them as factors in the infinite-coupling 
limits of Lagrangian field theories.  
The requirements of matching simple features of the low energy 
effective actions, global symmetry groups, and current algebra 
central charges, turn out to tightly constrain the possible 
properties of the SCFTs.  
In this paper we use this approach to map out rank 1 SCFTs (those 
with 1 complex-dimensional Coulomb branches) and find three new 
such theories.  

The singular Seiberg-Witten curves describing the low energy 
effective actions
on the Coulomb branches of rank 1 $N=2$ SCFTs were found in
\cite{apsw9511,mn9608} and correspond to Kodaira's classification 
\cite{kodaira} of degenerations of one-dimensional families of 
elliptic curves.
This gives a list of singularities associated to Lagrangian conformal
theories, together with six strongly-coupled isolated fixed point
singularities, three of which are characterized by having Coulomb 
branch vevs of dimensions 3, 4, and 6.   
Mass deformations of these singularities consistent with flavor 
symmetries $E_6$, $E_7$, and $E_8$ were constructed by Minahan and 
Nemeschansky \cite{mn9608}, and their existence was deduced from 
string constructions \cite{excft}.

We will argue on the basis of the proposed infinite coupling duality 
that there are other, inequivalent, mass deformations of these same 
singular curves, with correspondingly different flavor groups.  
In weakly-coupled examples, where we have a Lagrangian description, 
such inequivalent mass deformations are familiar: they correspond
to different choices of gauge representations for the matter
fields such that the gauge coupling remains marginal.
Our interacting fixed-point examples with the same singularity but 
different mass deformations can, heuristically, be thought of in the 
same way: they are strongly coupled rank 1 gauge theories all with 
the same ``gauge group" (corresponding to the singularity) but with 
different ``matter content" (corresponding to the different mass 
deformations).

Because we have different deformations of the same Coulomb
branch singularity, it is convenient to name the singularity
by the dimensions of its Coulomb branch vevs instead of by its 
global symmetry algebra, which is a property of a particular 
deformation of the singularity.  
Thus $\SCFT[d_i]$ will denote a rank-$n$ superconformal fixed 
point singularity with vevs of dimensions $d_1,\ldots,d_n$, and 
a particular deformation of this singularity with flavor symmetry 
algebra $\hf$ (corresponding to a particular SCFT) will be denoted 
by $\SCFT[d_i:\hf]$.
We denote a Lagrangian gauge theory with gauge algebra $\gf$ and 
half-hypermultiplet representation content $\br$ by $\gf\with \br$, 
from which the Coulomb branch vev dimensions and the flavor algebra
can be determined.\footnote{The rules: 
The Coulomb branch vevs are the orders of the independent adjoint 
Casimirs of $\gf$, which are the exponents of $\gf$ plus 1.  
The flavor symmetry depends on whether the half-hypermultiplets
are in complex, real (orthogonal), or pseudoreal (symplectic)
representations.
A complex representation always appears with its complex
conjugate, and $n\cdot(\br\oplus\bar\br)$ contributes a
$\U(n)$ flavor symmetry factor.  
Only even numbers of half-hypermultiplets in real representations 
can be coupled, and $2n\cdot\br$ contributes an $\Sp(n)$ flavor 
symmetry factor.  
Any number of pseudoreal half-hypermultiplets can be coupled, and 
$n\cdot\br$ gives an $\SO(n)$ flavor symmetry factor.}
In this notation, the duality proposed in \cite{as0711} has the 
general form
\be\label{icd}
\gf[d_i]\with\br = \tgf[\til d_i]\with(\tbr\oplus\SCFT[d:\hf]),
\ee
where we have indicated the dimensions of the Coulomb branch vevs 
of the Lagrangian gauge groups as well, for clarity.

\TABLE[ht]
{\begin{tabular}{|c|c|c|c|c|c|} \hline
$d$   & $\hf$   & $k_\hf$  &$\tf32\cdot k_R$&$48\cdot a$& $\Z_2$ obstruction? \\
\hline\hline
6         & $E_8$    & 12      & 124               & 190       & no      \\
6         & $\Sp(5)$ & 7       & 98                & 164       & yes     \\ 
4         & $E_7$    & 8       & 76                & 118       & no      \\
4& $\Sp(3)\oplus\SU(2)$& $5\oplus8$ & 58           & 100 &yes $\oplus$ no\\ 
3         & $E_6$    & 6       & 52                &  82       & no      \\
3         & $2\le\mbox{rank}(\hf)\le6$    
& $\le$ 8    & 34--38            &  64--68   & ?       \\ 
\hline
\end{tabular}
\caption{Properties of predicted rank 1 SCFTs.  $d$ is the
dimension of the Coulomb branch vev, $\hf$ is the flavor symmetry
algebra, $k_\hf$ is the flavor current algebra central charge,
$k_R$ is the $\U(1)_R$ current algebra central charge, and $a$ is one
of the conformal anomalies.  Only ranges of possible values are given
for the entry in the last row.\label{tab1}}}
Our main results are the existence and properties of
the rank one isolated SCFTs shown in table 1.
The curves of the $E_{6,7,8}$ theories were found in \cite{mn9608}.
The central charges of the $E_6$ and $E_7$ theories were found by
consistency of the infinite-coupling duality proposal in 
\cite{as0711}.   In the rest of this note we use 
similar consistency arguments to compute the other entries in
table 1.  These properties agree with those of the $E_{6,7,8}$ 
theories computed using AdS/CFT techniques in \cite{at0711}.

The evidence for the new SCFTs comes from finding
many different examples of dualities of the form (\ref{icd}) with 
SCFT factors with the properties shown in table 1.  As we
describe in more detail below, we search through a list
of Lagrangian conformal theories with the assumption that
they have infinite coupling duals of the form (\ref{icd}),
and compute the properties of the assumed isolated SCFT
by matching various symmetries and anomalies on the two
sides of the duality.  Whenever we find two or more examples
giving matching SCFTs, we include it
in table 1.  In fact, the resulting set of dualities gives
many examples for each of the theories, consistent
in an intricate and beautiful way, and are listed in tables 2
and 3 below.
The exception is the SCFT in the last line of table 1 for which
there is only one duality which does not give enough constraints
to pin down its properties precisely.
In an appendix we include some notes on computing Lie algebra
embeddings, and record the embeddings used in tables 2 and 3,
for the convenience of readers interested in checking our results.


Constraints on possible infinite coupling 
duals of the form (\ref{icd}) come from matching on both sides
of the equivalence the following seven quantities:
\begin{enumerate}
\item The rank of the gauge group and the spectrum of dimensions of 
the Coulomb branch vevs, implying
\be\label{rank}
\{d_i\} = \{\til d_i\}\cup\{d\} .
\ee
\item The flavor symmetry algebras, implying the flavor symmetry 
on the left is the sum, $\tff\oplus\ff$, of the flavor symmetry 
$\tff$ of the $\tbr$ half-hypermultiplets on the  right and
the commutant of $\tgf$ in $\hf$, 
\be\label{embed}
\tgf\oplus\ff\subset \hf\ \ \ \ \mbox{with $\ \ff\ $ maximal.}
\ee
\item The contribution to the beta function from weakly gauging the
flavor symmetry on both sides, giving
\be\label{beta}
T(\br) = T(\tbr) + k_{\hf} \cdot I_{\ff\hra\hf},
\ee
where $T(\br)$ is the quadratic index of the representation
$\br$, $I_{\ff\hra\hf}$ is the Dynkin index of embedding, and
$k_\hf$ is the central 
charge of the $\hf$ flavor symmetry current algebra.
\item The number of marginal couplings, implying the vanishing of the
beta function of the $\tgf$ gauge factor,
\be\label{k}
2\cdot T(\tgf) = T(\tbr) + k_{\hf} \cdot I_{\tgf\hra\hf},
\ee
where $T(\tgf)$ denotes the quadratic index of the adjoint representation of $\tgf$. 
\item The contribution to the $\U(1)_R$ symmetry central charge on 
both sides, giving this central charge of the SCFT, $k_R$, as
\be\label{c}
(3/2)\cdot k_R = 24\cdot c
= 4\cdot (|\gf|-|\tgf|) + (|\br|-|\tbr|),
\ee
which is related as shown to the energy-momentum central charge $c$.
$|\gf|$ and $|\br|$ are the dimensions of the algebra and
representation, respectively.
\item The contribution to the $a$ conformal anomaly, giving
$a$ for the SCFT as
\be\label{a}
48\cdot a = 10\cdot (|\gf|-|\tgf|) + (|\br|-|\tbr|).
\ee
In general $|\gf|=\sum_i (2d_i-1)$, 
implying from (\ref{rank}) that $|\gf| - |\tgf| = 2d-1$.  So
the difference between (\ref{c}) and (\ref{a})
is already fixed by condition i.
\item Whether there is a global $\Z_2$ obstruction \cite{w82} 
to gauging of the flavor symmetry.
\end{enumerate}
The first five conditions were described in \cite{as0711},
while the $a$ and $c$ conformal anomalies can be computed
from 't Hooft anomaly matching, as reviewed in \cite{at0711}.
The global $\Z_2$ obstruction matching is described below.
Our conventions for the normalization of the central
charges and for the quadratic index follow those of \cite{as0711}.

Let us illustrate the use of these constraints in determining
the dual SCFT in the case of the the Lagrangian
$G_2\with 8\cdot\bf7$ conformal theory, given as entry 4 in table 2. 
We assume it has an infinite-coupling dual of the form
(\ref{icd}).  Apply the above list of conditions:
\begin{enumerate}
\item 
Since $G_2$ is rank 2, with adjoint Casimirs
of order 2 and 6,
the only possibility is that $\SCFT[\hf]$ and $\tgf$ are both rank 1 gauge
theories.  Thus we must have $\tgf=\SU(2)$, and since its Coulomb branch 
vev has dimension 2, the dimension 6 vev must belong to the rank 1 
$\SCFT[6:\hf]$.\footnote{In this particular case, though not in
most of the other cases we consider, there is strong additional
evidence for these conclusions:  the form of the curve describing
the low energy effective action of the $G_2$ theory is known \cite{acsw0504}, 
from which it is easily checked that it factorizes into
the scale invariant $\SU(2)$ singularity \cite{sw9408} and the
$\SCFT[6]$ singularity in the infinite coupling limit.}
Thus
\be\label{g2w8.7}
G_2\with 8\cdot{\bf7} = \SU(2) \with ( n\cdot{\bf2}\oplus \SCFT[6:\hf] )
\ee
with $\SU(2) \subset \hf$.
Only $n\le 7$ half-hypermultiplet $\bf2$'s can occur since any more or
any other representations would contribute too much to the $\SU(2)$
beta function, making it infrared free.  
\item
The left side of (\ref{g2w8.7}) has $\Sp(4)$ flavor symmetry since 
the $\bf7$ is a real representation.  Since the $\bf2$ of $\SU(2)$ 
is pseudoreal, the half-hypermultiplets on the right side contribute
an $\SO(n)$ flavor symmetry factor.  Thus the only way to match the global
symmetries on both sides is for $n=0$ or $1$ (so that the half-hypermultiplets
don't contribute any flavor factors), and $\SU(2) \oplus \Sp(4) 
\subset \hf$ with $\Sp(4)$ being the commutant of $\SU(2)$ in $\hf$.
Assuming that $\hf$ is simple and has rank less than or equal to 8, 
some work with a table of maximal subalgebras \cite{mp81} shows that
only $\SO(16)$ and $\Sp(5)$ have $\SU(2)$ subalgebras with
commutant $\Sp(4)$.  In either case the $\Sp(4)$ factor has Dynkin
index of embedding $I_{\Sp(4)\hra\SO(16)}=I_{\Sp(4)\hra\Sp(5)}=1$,
but $I_{\SU(2)\hra\SO(16)}=4$, while $I_{\SU(2)\hra\Sp(5)}=1$.
\item For the $\Sp(4)$ flavor symmetry
$7 = 7\cdot T({\bf8})= k_\hf I_{\Sp(4)\hra\hf}$.  (The
$n\cdot{\bf2}$'s don't contribute because they are singlets 
under the $\Sp(4)$.)
Since the index of embedding is 1, we find that $k_\hf=7$ whether
$\hf=\SO(16)$ or $\Sp(5)$.
\item
$8=T(\SU(2))=n+7\cdot I_{\SU(2)\hra\hf}$.
Since $n\in\{0,1\}$ and the index of embedding is a positive integer, the
only solution is $n=1$ and $I_{\SU(2)\hra\hf}=1$.  Thus we must have
$\hf=\Sp(5)$.
\item
$(3/2)\cdot k_R = 4\cdot(14-3) + (7\cdot8-1\cdot2) = 98$.
\item
$48\cdot a = 10\cdot(14-3)+(7\cdot8-1\cdot2) = 164$.
\item
The $\Sp(4)$ flavor symmetry of $G_2\with 8\cdot\bf7$ has
a global $\Z_2$ obstruction to being gauged since there is 
an odd number (7) of the pseudoreal $\bf8$'s of $\Sp(4)$.  
So, to weakly gauge
the $\Sp(4)$, a spectator half-hypermultiplet (or just a 
Weyl fermion) in a pseudoreal representation must be added.
Since the spectator fields are otherwise uncoupled, this
$\Z_2$ obstruction must persist whatever the value of the $G_2$ 
coupling, and so should also be seen in the dual $\SU(2)
\with ( {\bf2}\oplus\SCFT[6:\Sp(5)] )$ theory.
Since the SCFT factor contributes the fields transforming
under $\Sp(4)$, and since $I_{\Sp(4)\hra\Sp(5)}=1$ is odd,
it follows that the $\SCFT^{[6]}_{\Sp(5)}$ theory by itself
must have a global $\Z_2$ obstruction in its $\Sp(5)$ flavor
algebra.\footnote{Thanks to N. Seiberg for discussions on 
this point.}  This is necessary also for the $\SU(2)$ gauge
factor to be anomaly-free: the single $\bf2$ half-hypermultiplet
contributes a $\Z_2$ anomaly, but so does the SCFT factor 
since $\SU(2)$ is of odd index in $\Sp(5)$.
\end{enumerate}
We assumed in the above argument that rank$(\hf)\le8$ 
and $\hf$ was simple.  The first assumption is justified because 
the maximum number of independent mass deformations of 
$\SCFT[6]$ is 8---as seen by counting the independent 
deformations of the complex structure of its curve---so whatever 
its flavor algebra it must have rank at most 8.
The simplicity of $\hf$ assumption has no justification, and
without it there are many more possible answers.  For example, 
we could have $\hf=\SU(2)\oplus\Sp(4)$.  In these non-simple
cases, each simple factor can have a different central
charge, making it easy to satisfy the requirements
to be a dual description.

Even though each individual assumed duality may be consistent
with more than one possible isolated rank 1 SCFT, we can
gain stronger evidence for the existence of a particular one
by showing that it consistently occurs in the duals of
other theories.  We can check for this by examining 
higher-rank Lagrangian conformal theories which could
have infinite coupling duals with $\SCFT[d]$ factors with
$d=3$, $4$, or $6$.  The possible Lagrangian theories are
constrained by requirement (\ref{rank}) above.  For example, for the 
infinite-coupling dual to contain a $\SCFT[6]$ factor, the 
gauge algebra $\gf$ of the Lagrangian theory must include an 
order 6 adjoint Casimir, and its list of remaining Casimir 
orders must be those of the dual gauge algebra $\tgf$.  There are only 
a handful of
possibilities for such $\gf$ among simple algebras, namely
$\SU(3)$, $\Sp(2)$, $G_2$, $\SU(4)$, $\SO(7)$, $\Sp(3)$, $\SO(8)$, 
$\SU(6)$, and $\SO(12)$.  From the known curves for the rank 2
theories, only 6 have infinite-coupling limits \cite{as0711}, and
there are 53 (non-$N{=}4$) Lagrangian conformal theories with gauge 
algebras, $\gf$, with rank $\ge3$ in this list.  By searching through 
this list for pairs of theories consistent with duals involving 
the same SCFTs, we find the 16 ones shown
in table 2, all consistent with the properties recorded in table 1.
Some details of the flavor symmetry 
embeddings for each entry in table 2 are collected in the appendix.
\TABLE[ht]
{{
\begin{tabular}{|r|ll@{$\ \ =\ \ $}lll|}\hline
&\multicolumn{1}{l@{$\ \with$}}{$\gf$}&$\br$&
\multicolumn{1}{l@{$\ \with$}}{$\!\!\!\tgf\ \ $}&
\multicolumn{1}{l@{$\SCFT\!\!\!$}}{$\tbr\qquad\oplus$}&
$[d:\hf]$\\ \hline\hline
1.&
$\Sp(3)$ & ${\bf14}\oplus11\cdot{\bf6}$ &
$\Sp(2)$ & & $[6:E_8]$ \\
2.&
$\SU(6)$ & ${\bf20}\oplus{\bf15}\oplus\bar{\bf15}\oplus
5\cdot{\bf6}\oplus5\cdot\bar{\bf6}$ &
$\SU(5)$ &  ${\bf5}\oplus\bar{\bf5}\oplus
{\bf10}\oplus\bar{\bf10}$ & $[6:E_8]$ \\ 
3.&
$\SO(12)$ & $3\cdot{\bf32}\oplus{\bf32}'\oplus4\cdot{\bf12}$ &
$\SO(11)$ & $3\cdot{\bf32}$ & $[6:E_8]$ \\
4.&
$G_2$ & $8\cdot{\bf7}$ &
$\SU(2)$ & ${\bf2}$ & $[6:\Sp(5)]$ \\
5.&
$\SO(7)$ & $4\cdot{\bf8}\oplus6\cdot{\bf7}$ &
$\Sp(2)$ & $5\cdot{\bf4}$ & $[6:\Sp(5)]$ \\
6.&
$\SU(6)$ & ${\bf21}\oplus\bar{\bf21}\oplus{\bf20}
\oplus{\bf6}\oplus\bar{\bf6}$ &
$\SU(5)$ & ${\bf10}\oplus\bar{\bf10}$ & $[6:\Sp(5)]$ \\ 
7.&
$\Sp(2)$ & $12\cdot{\bf 4}$ & 
$\SU(2)$ &\multicolumn{1}{c}{} & $[4:E_7]$\\
8.&
$\SU(4)$ & $2\cdot{\bf6}\oplus6\cdot{\bf4}\oplus6\cdot\bar{\bf4}$ &
$\SU(3)$ & $2\cdot{\bf3}\oplus2\cdot\bar{\bf3}$ & $[4:E_7]$ \\
9.&
$\SO(7)$ & $6\cdot{\bf8}\oplus4\cdot{\bf7}$ &
$G_2$ &  $4\cdot{\bf7}$ & $[4:E_7]$ \\
10.&
$\SO(8)$ & $6\cdot{\bf8}\oplus4\cdot{\bf8}'
\oplus2\cdot{\bf8}''$ &
$\SO(7)$ & $6\cdot{\bf8}$ & $[4:E_7]$ \\
11.&
$\SO(8)$ & $6\cdot{\bf8}\oplus6\cdot{\bf8}'$ &
$G_2$ &  & $
[4:E_7] \oplus [4:E_7]$ \\ 
12.&
$\Sp(2)$ & $6\cdot{\bf5}$ &
$\SU(2)$ & & $[4:\Sp(3)\oplus\SU(2)]$ \\
13.&
$\Sp(2)$ & $4\cdot{\bf4}\oplus4\cdot{\bf5}$ &
$\SU(2)$ & $3\cdot{\bf2}$& $[4:\Sp(3)\oplus\SU(2)]$ \\
14.&
$\SU(4)$ & ${\bf10}\oplus\bar{\bf10}\oplus2\cdot{\bf4}\oplus2\cdot\bar{\bf4}$&
$\SU(3)$ & ${\bf3}\oplus\bar{\bf3}$ & $[4:\Sp(3)\oplus\SU(2)]$ \\
15.&
$\SU(3)$ & $6\cdot{\bf3}\oplus6\cdot\bar{\bf3}$ & 
$\SU(2)$ & $2\cdot{\bf 2}$ & $[3:E_6]$  \\
16.&
$\SU(4)$ & $4\cdot{\bf6}\oplus4\cdot{\bf4}\oplus4\cdot\bar{\bf4}$ &
$\Sp(2)$ & $6\cdot{\bf4}$ & $[3:E_6]$ \\ 
17.&
$\SU(3)$ & ${\bf3}\oplus\bar{\bf3}\oplus{\bf6}\oplus\bar{\bf6}$ &
$\SU(2)$ & $n\cdot{\bf2}$ & $[3:\hf]$ \\ 
\hline
\end{tabular}
}
\caption{Predicted dualities with one marginal operator.\label{tab2}}}

That not all of the 53 rank $\ge3$ Lagrangian theories appear 
in table 2 is
reasonable, since they need not all necessarily
have rank 1 SCFT factors: they could be self-dual, or 
could be dual to theories with rank $\ge$2 SCFT factors.  
For example, from the known curves for the Lagrangian
superconformal theories with rank $r$ classical gauge groups with
hypermultiplets in fundamental representations \cite{as9509},
it follows from their degeneration in the infinite coupling 
limit that their duals are $\SU(2)$ plus rank $r-1$ isolated
SCFTs.

Entry 17 of table 2 is the only case where we do not have enough
information to determine the properties of the dual SCFT
precisely.  Because the flavor symmetry is $\U(1)\oplus\U(1)$,
we can only have $n\in\{0,1,2\}$, and must have $\SU(2)_I \oplus
\U(1) \subset \hf$, with $\U(1)$ maximal and $I$ the index of embedding
of the $\SU(2)$.  Then $k_{\hf}=(8-n)/I$, $(3/2)\cdot k_R 
= 38-2n$, and $48\cdot a=68-2n$.  There are many possibilities
for $\hf$ consistent with the embedding constraint.  
Nevertheless, since $k_R$ and $a$ are
smaller than those of the $\SCFT[3:E_6]$ theory, and since
it is known from its curve \cite{lll9805} that the $\SU(3)\with
\bf3\oplus\bar3\oplus6\oplus\bar6$ theory has an infinite coupling 
limit, it follows that there must be a new $\SCFT[3]$.

There is another, less direct, way of constraining $\hf$ for
entry 17.  We can reverse the direction of our logic, and,
starting with $\SCFT[3:\hf]$ we can gauge different $\SU(2)$ 
subalgebras of $\hf$ and try to add appropriate numbers of doublet
half-hypermultiplets to make the gauge coupling marginal.
By construction there will be one such embedding giving the
Lagrangian conformal theory of entry 17.  But if there
are other embeddings, then we would predict the existence
of a rank 2 conformal theory with Coulomb branch dimensions
$2$ and $3$ and with a marginal coupling, but no purely
weakly coupled (Lagrangian) limit.  There is nothing wrong
with this in general---indeed, their existence for higher
ranks is a robust prediction of infinite-coupling duality, 
as we will discuss below.  However, at rank 2 there is some
evidence from systematic searches for all possible curves
of rank 2 SCFTs \cite{acsw0504,aw0510} that all those curves
with marginal operators have a limit in the coupling space
where the curve becomes singular in a way consistent with 
a purely weakly coupled Lagrangian description.  This is
further supported by the fact that for the other 5 rank 1
SCFTs listed in table 1, the only conformal gaugings of 
$\SU(2)$ subalgebras are precisely those with Lagrangian
limits.  If we assume then that this should also apply to
the $\SCFT[3:\hf]$ theory, it then follows from a
detailed examination of algebra embeddings that rank$(\hf)=2$,
since all higher-rank $\hf$ turn out to admit multiple
inequivalent conformal gaugings of $\SU(2)$ subalgebras.
Assuming rank$(\hf)=2$ forces $n=2$, so gives $k_\hf=6/I$,
$(3/2)\cdot k_R= 34$ and $48\cdot a=64$, but does not 
constrain $\hf$ any further since all rank 2 $\hf$'s only
admit the single Lagrangian conformal $\SU(2)$ embedding.

In general there are many different ways of gauging rank $\ge2$ 
subalgebras of the flavor algebras of the SCFTs in table 1.  This
provides many examples of conformal theories with marginal couplings
which do not appear in the list in table 2 of such theories with
purely Lagrangian descriptions in some limit.  Some simple 
examples are 
\be\label{nonLaglist}
\begin{array}{lcl}
\SU(3) \with \SCFT[3:E_6] &\qquad&
E_6 \supset\SU(3)_2\oplus (G_2)_1 ,\\
\SU(3) \with \SCFT[6:E_8] &\qquad&
E_8 \supset\SU(3)_1\oplus (E_6)_1 ,\\
G_2 \with 2\cdot{\bf7}\oplus\SCFT[6:E_8] &&
E_8 \supset (G_2)_1 \oplus (F_4)_1 ,\\
F_4 \with 4\cdot{\bf26}\oplus\SCFT[6:E_8] &&
E_8 \supset (F_4)_1 \oplus (G_2)_1 ,
\end{array}
\ee
where we have indicated the embedding of the gauge algebra 
in the SCFT flavor algebra on the right, with the index of 
embedding of each subalgebra shown as a subscript.  For
these examples it is clear that they cannot be dual to
a purely Lagrangian field theory, since they all have
exceptional flavor algebras.  These are thus examples of
conformal theories with marginal couplings but no purely
weak-coupling limit.

\TABLE[ht]
{{\small
\begin{tabular}{|r|ll@{$\ \ =\ \ $}ll|}\hline
&\multicolumn{1}{l@{$\with$}}{$\gf$}&$\br$&
\multicolumn{1}{l@{$\with$}}{$\!\!\!\tgf\ \ $}&
$\tbr\hfill\oplus\hfill\SCFT[d:\hf]$\\ \hline\hline
18.&
$\SU(2)\oplus\SU(3)$ & $2\cd({\bf2},{\bf1})\oplus({\bf2},{\bf3}\op
\bar{\bf3})\oplus4\cd({\bf1},{\bf3}\op\bar{\bf3})$ &
$\SU(2)\oplus\SU(2)$ & $2\cd({\bf2},{\bf1})\oplus2\cd({\bf1},{\bf2})
\hfill
[3:E_6]$ \\
19.&
$\SU(2)\oplus\Sp(2)$ & $2\cd({\bf2},{\bf4})\oplus8\cd({\bf1},{\bf4})$ &
$\SU(2)\oplus\SU(2)$ & \hfill  $[4:E_7]$ \\ 
20.&
$\SU(2)\oplus\Sp(2)$ & $3\cd({\bf2},{\bf1})\oplus({\bf2},{\bf5})
\oplus4\cd({\bf1},{\bf5})$ &
$\SU(2)\oplus\SU(2)$ & $3\cd({\bf2},{\bf1})
\hfill
[4:\Sp(3)\op\SU(2)]$ \\
21.&
$\SU(2)\oplus G_2$ & $({\bf2},{\bf1})\oplus({\bf2},{\bf7})
\oplus6\cd({\bf1},{\bf7})$ & 
$\SU(2)\oplus\SU(2)$ & $({\bf2},{\bf1})\oplus({\bf1},{\bf2})
\hfill
[6:\Sp(5)]$ \\
\hline
22.&
$\SU(3)\oplus\SU(3)$ & $2\cd({\bf3},\bar{\bf3})\oplus
2\cd(\bar{\bf3},{\bf3})$ &
$\SU(2)\oplus\SU(3)$ & $2\cd({\bf2},{\bf1})
\hfill
[3:E_6]$ \\
23.&
$\SU(3)\oplus\SU(3)$ & $({\bf3}\op\bar{\bf3},{\bf3}\op\bar{\bf3})$ &
$\SU(2)\oplus\SU(3)$ & $2\cd({\bf2},{\bf1})
\hfill
[3:E_6]$ \\ 
24.&
$\SU(3)\oplus\SU(3)$ & $3\cd({\bf3}\op\bar{\bf3},{\bf1})\oplus
({\bf3},\bar{\bf3})\oplus(\bar{\bf3},{\bf3})$
&$\SU(2)\oplus\SU(3)$ &$2\cd({\bf2},{\bf1})
\hfill
[3:E_6]$\\
&&\multicolumn{1}{l}{$\qquad\qquad\qquad\qquad \mbox{}\oplus
3\cd({\bf1},{\bf3}\op\bar{\bf3})$}
&&$\quad \mbox{}\oplus3\cd({\bf1},{\bf3}\op\bar{\bf3})$
\\
25.&
$\SU(3)\oplus\Sp(2)$ & $({\bf3}\op\bar{\bf3},{\bf1})\oplus
({\bf3}\op\bar{\bf3},{\bf5})$ 
&$\SU(2)\oplus\Sp(2)$ & $2\cd({\bf2},{\bf1})
\hfill
[3:E_6]$ \\
&&&
$\SU(3)\oplus\SU(2)$ &  $({\bf3}\op\bar{\bf3},{\bf1})
\hfill
[4:\Sp(3)\op\SU(2)]$ \\
26.&
$\SU(3)\oplus\Sp(2)$ & $2\cd({\bf3}\op\bar{\bf3},{\bf1})\oplus
({\bf3}\op\bar{\bf3},{\bf4})\oplus6\cd({\bf1},{\bf4})$ &
$\SU(2)\oplus\Sp(2)$ & $2\cd({\bf2},{\bf1})\oplus6\cd({\bf1},{\bf4})
\hfill
[3:E_6]$ \\
&&&
$\SU(3)\oplus\SU(2)$ & $2\cd({\bf3}\op\bar{\bf3},{\bf1})
\hfill
[4:E_7]$ \\ 
27.&
$\Sp(2)\oplus\Sp(2)$ & $2\cd({\bf5},{\bf1})\oplus({\bf5},{\bf4})\oplus
7\cd({\bf1},{\bf4})$ &
$\SU(2)\oplus\Sp(2)$& $7\cd({\bf1},{\bf4})
\hfill
[4:\Sp(3)\op\SU(2)]$ \\
&&&
$\Sp(2)\oplus\SU(2)$ & $2\cd({\bf5},{\bf1})
\hfill
[4:E_7]$ \\
28.&
$\Sp(2)\oplus\Sp(2)$ & $4\cd({\bf4},{\bf1})\oplus2\cd({\bf4},{\bf4})\oplus
4\cd({\bf1},{\bf4})$ & 
$\SU(2)\oplus\Sp(2)$ & $4\cd({\bf1},{\bf4})
\hfill
[4:E_7]$  \\
29.&
$\Sp(2)\oplus G_2$ & $5\cd({\bf4},{\bf1})\oplus({\bf4},{\bf7})\oplus
4\cd({\bf1},{\bf7})$ &
$\SU(2)\oplus G_2$ & $4\cd({\bf1},{\bf7})
\hfill
[4:E_7]$ \\ 
&&&
$\Sp(2)\oplus\SU(2)$ & $5\cd({\bf4},{\bf1})\oplus({\bf1},{\bf2})
\hfill
[6:\Sp(5)]$ \\ 
\hline
\end{tabular}
}
\caption{Some predicted dualities with two marginal operators.\label{tab2}}}
All our arguments can also be applied to theories with
more than one marginal gauge coupling.  We illustrate this on the
small set of rank 2 and 3 Lagrangian theories with two marginal couplings
which reduce in the limit as one of the couplings becomes weak to
one of the rank 2 theories (entries 4, 7, 12, 13, 15, or 17) in 
table 2.
In the rank 3 cases, one factor of the gauge algebra must be $\SU(2)$
which is self dual.  The dual descriptions in the infinite-coupling limit
of the other factor's coupling are given in entries 18 to 21 of
table 3.  For the rank 4 examples, where each simple gauge algebra
factor has rank 2, we give their dual descriptions in table 3 only
in the limit as one or the other gauge factor is taken to infinite
coupling.  (If the two infinite coupling limits are different,
we give the dual of the first factor on the first line and that of the
second factor on the second line of the entry in table 3.)
The double infinite-coupling limit of all these theories have
dual descriptions of the form $\SU(2)\oplus\SU(2)\with{\br}\oplus
\SCFT[d_1,d_2]$ where $\SCFT[d_1,d_2]$ are isolated rank 2 SCFTs.
The success of the infinite coupling dual descriptions of all these 
theories is strong additional evidence for the existence of the 
first five SCFTs in table 1.

The existence of the new rank 1 SCFTs with flavor symmetries
other than $E_n$ raises some obvious questions:  What are
the Seiberg-Witten curves and one-forms describing the low energy
effective actions of (the mass deformations of) these theories?
Are there string constructions that realize these sub-maximal
SCFTs?  

The set of dual descriptions found in tables 2 and 3, together
with the known Seiberg-Witten singular curves for the rank 1
SCFTs and some Lagrangian conformal theories that appear on the
right side of the duality can be used in many
cases to determine the singular curves for the higher-rank 
Lagrangian theories on the left.

The techniques of this paper can also be applied to determining
the properties of higher-rank isolated SCFTs.  Here, though, the
story will certainly be much more complicated:  there are many
known rank 2 singularities \cite{acsw0504,aw0510}, a complete
list has not been found, and techniques for determining their
mass deformations consistent with the requirements of $N=2$
supersymmetry have not been developed.

\section*{Acknowledgments}
It is a pleasure to thank 
O. Aharony,
F.P. Esposito,
N. Seiberg,
A. Shapere,
and R. Wijewardhana 
for helpful comments and discussions.  
This work is supported in part by DOE grant FG02-84-ER40153. 

\appendix
\section*{Appendix}






The embeddings $\tgf\subset\hf$ for each of the dual theories
listed in tables 2 and 3 are listed below in table 4, where the
Dynkin index of embedding of each subalgebra is shown as a
subscript.
We make a few comments on how these embeddings can be extracted
from tables.  
Tables of maximal semisimple subalgebras of simple Lie 
algebras are given in \cite{mp81}.  We need not only the
semisimple factors, but also any $\U(1)$ factors that may
occur.  Maximal reductive subalgebras of simple Lie algebras
which have an abelian factor only have a single $\U(1)$, and
their semisimple factors correspond to the Dynkin diagram
which results from eliminating any two nodes with mark 1 from
the extended Dynkin diagram of the original algebra \cite{kmps90}.

All embeddings can be found by following chains of maximal
embeddings.  One is then faced with finding maximal embeddings
in semi-simple algebras.  The only non-trivial case then are
diagonal embeddings in two or more identical factors.  For
example, $\hf\supset \gf_J\oplus\gf_K\supset \gf_{J+K}\oplus
\U(1)\oplus\cdots\oplus\U(1)$ where there are rank$(\gf)$
$\U(1)$ factors; the subscripts are to indicate that the
Dynkin indices of embedding add under diagonal embedding.

The problem we face is not just to find an embedding of a
gauge algebra $\tgf$ in $\hf$, but also to compute the commutant
of $\tgf$ in $\hf$.  In general, upon following a chain of
maximal embeddings, the resulting subalgebra commuting with
$\tgf$ need not be maximal.  However, the maximal commuting 
subalgebra will appear somewhere in the tree of all possible 
chains of maximal embeddings which contain $\tgf$ as a factor.
To determine whether a given commuting subalgebra is maximal 
or is itself a subalgebra of a commuting subalgebra found
in a different chain, one needs to check whether the two 
different embeddings of $\tgf$ are equivalent or not.

For example, consider the case of entries 22 and 23 in table 3.
These are two different Lagrangian theories whose duals have
the same gauge group $\tgf=\SU(2)_1\oplus\SU(3)_2$ both embedded 
in the flavor algebra $\hf=E_6$ of $\SCFT[3:E_6]$.  (Theory 22
is an ``elliptic model" whose low energy effective action
was computed in \cite{w9703}, while theory 23 is a ``twisted
elliptic" model whose curve is not known.)
In the case of theory 22 the chain of maximal embeddings is
\be\label{22embed}
E_6 \supset \SU(3)_2 \oplus \biggl( 
(G_2)_1 \supset \SU(2)_1 \oplus \SU(2)_3 \biggr),
\ee
while for theory 23 it is
\be\label{23embed}
E_6 \supset \SU(2)_1 \oplus \biggl( 
\SU(5)_1 \supset \SU(4)_2 \supset \SU(3)_2 \oplus \U(1) \biggr).
\ee
To show that these are really different maximal embeddings, we
must show that they are different embeddings of $\SU(2)_1\oplus
\SU(3)_2$ in $E_6$, otherwise (\ref{23embed}) would just be
a subalgebra of (\ref{22embed}).
To show this, we check the branching of a specific representation.
For (\ref{22embed}), under $E_6 \supset \SU(3) \oplus G_2$: 
$\bf27=(\bar6,1) \oplus(3,7)$, 
and under $G_2 \supset \SU(2)_1 \oplus \SU(2)_3$: 
$\bf7=(2,2)\oplus(1,3)$.  
For (\ref{23embed}), under $E_6 \supset \SU(2)_1 \oplus \SU(5)_1$: 
$\bf27=(1,\bar{15})\oplus(2,6)$,
and under $\SU(6)\supset\SU(3)\supset\SU(2)\oplus\U(1)$:
$\bf\bar{15}=15=8\oplus3\oplus\bar3\oplus1$ and $\bf6=6=3\oplus\bar3$.
Putting these together, under the (\ref{22embed}) embedding
$$
E_6\supset\SU(3)_2\oplus\SU(2)_1\ \ :\ \ {\bf27 = (\bar6,1)}\oplus
2\cdot({\bf3,2})\oplus3\cdot(\bf{3,1}),
$$ 
while under the (\ref{23embed}) embedding 
$$
E_6\supset\SU(3)_2\oplus\SU(2)_1\ \ :\ \ \bf27 = (8,1)\oplus(3,1)\oplus
(\bar3,1)\oplus(1,1)\oplus(3,2)\oplus(\bar3,2),
$$
showing that they are inequivalent embeddings.

\TABLE[ht]
{{
\begin{tabular}[ht]{|r|l@{$\ \ \supset\ \ $}l@{$\ \ \oplus\ \ $}l|}\hline
&$\hf$&$\tgf_{I}$&$\ff_{I'}$\\ \hline\hline
1.&
$E_8$& $\Sp(2)_1$& $\SO(11)_1$\\
2.&
$E_8$& $\SU(5)_1$& $\SU(5)_1$\\
3.&
$E_8$& $\SO(11)_1$& $\Sp(2)_1$\\
4.&
$\Sp(5)$& $\SU(2)_1$& $\Sp(4)_1$\\
5.&
$\Sp(5)$& $\Sp(2)_1$& $\Sp(3)_1$\\
6.&
$\Sp(5)$& $\SU(5)_2$& $\U(1)$\\
7.&
$E_7$& $\SU(2)_1$& $\SO(12)_1$\\
8.&
$E_7$& $\SU(3)_1$& $\SU(6)_1$\\
9.&
$E_7$& $(G_2)_1$& $\Sp(3)_1$\\
10.&
$E_7$& $\SO(7)_1$& $\Sp(2)_1\oplus\SU(2)_1$\\
11.&
$E_7$& $(G_2)_1$& $\Sp(3)_1\qquad\ $ for both\\
12.&
$\Sp(3)\oplus\underline{\SU(2)}$& $\underline{\SU(2)_1}$& $\Sp(3)_1$\\
13.&
$\Sp(3)\oplus\underline{\SU(2)}$& $\SU(2)_1$
& $\Sp(2)_1\oplus\underline{\SU(2)_1}$\\
14.&
$\Sp(3)\oplus\underline{\SU(2)}$& $\SU(3)_2$
& $\U(1)\oplus\underline{\SU(2)_1}$\\
15.&
$E_6$& $\SU(2)_1$& $\SU(6)_1$\\
16.&
$E_6$& $\Sp(2)_1$& $\Sp(2)_1\oplus\U(1)$\\
17.&
$\hf$& $\SU(2)_I$& $\U(1) \qquad\quad\ $ if $n=2$\\
&& $\SU(2)_I$& $\U(1)\oplus\U(1)\ $ if $n=1,0$\\
\hline
18.&
$E_6$& $\SU(2)_1\oplus\SU(2)_1$& $\SU(4)_1$\\
19.&
$E_7$& $\SU(2)_1\oplus\SU(2)_1$& $\SU(2)_1\oplus\SO(8)_1$\\
20.&
$\Sp(3)$& $\SU(2)_1$& $\Sp(2)_1$\\
21.&
$\Sp(5)$& $\SU(2)_1\oplus\SU(2)_1$& $\Sp(3)_1$\\
\hline
22.&
$E_6$& $\SU(3)_2\oplus\SU(2)_1$& $\SU(2)_3$\\
23.&
$E_6$& $\SU(3)_2\oplus\SU(2)_1$& $\U(1)$\\
24.&
$E_6$& $\SU(3)_1\oplus\SU(2)_1$& $\SU(3)_1\oplus\U(1)$\\
25.&
$E_6$& $\SU(2)_1\oplus\Sp(2)_2$& $\U(1)$\\
&
$\Sp(3)\oplus\underline{\SU(2)}$& $\SU(3)_2\oplus\underline{\SU(2)_1}$
& $\U(1)$\\
26.&
$E_6$& $\SU(2)_1\oplus\Sp(2)_1$& $\SU(2)_1\oplus\U(1)$\\
&
$E_7$& $\SU(3)_1\oplus\SU(2)_1$& $\SU(4)_1\oplus\U(1)$\\
27.&
$\Sp(3)\oplus\underline{\SU(2)}$& $\underline{\SU(2)_1}\oplus\Sp(2)_1$
& $\SU(2)_1$\\
&
$E_7$& $\Sp(2)_1\oplus\SU(2)_1$& $\SO(7)_1$\\
28.&
$E_7$& $\Sp(2)_1\oplus\SU(2)_1$& $\SU(2)_1\oplus\SU(2)_1\oplus\SU(2)_2$\\
29.&
$E_7$& $(G_2)_1\oplus\SU(2)_1$& $\Sp(2)_1$\\
&
$\Sp(5)$& $\Sp(2)_1\oplus\SU(2)_1$& $\Sp(2)_1$\\
\hline
\end{tabular}
}
\caption{Algebra embeddings.  The subscript on each subalgebra
is its index of
embedding.\label{tab5}}}

\end{document}